\documentclass[intlimits,twoside,a4paper]{article}

\usepackage{amsmath,amssymb}
\usepackage{graphicx}

\usepackage[T2A]{fontenc}
\usepackage[cp1251]{inputenc}

\usepackage{cmpj3}

\issue{2017}{20}{4}{43706}
\doinumber{10.5488/CMP.20.43706}

\title[Energy spectrum of localized quasiparticles]%
{Energy spectrum of localized quasiparticles renormalized by multi-phonon processes at finite temperature}
\author[M.V.~Tkach, O.Yu.~Pytiuk, O.M.~Voitsekhivska, Ju.O.~Seti]{M.V.~Tkach\footnote{E-mail: ktf@chnu.edu.ua}\,,
O.Yu.~Pytiuk, O.M.~Voitsekhivska, Ju.O.~Seti}
\address{Chernivtsi National University, 2 Kotsyubinsky St.,
58012 Chernivtsi, Ukraine}

\date{Received July 12, 2017, in final form August 5, 2017}

\begin{document}
\maketitle

\begin{abstract}
The theory of renormalized energy spectrum of localized quasi-particle interacting with polarization phonons at finite temperature is developed within the Feynman-Pines diagram technique.
The created computer program effectively takes into account multi-phonon processes, exactly defining all diagrams of mass operator together with their analytical expressions in arbitrary order over the coupling constant. Now it is possible to separate the pole and non-pole mass operator terms and perform a partial summing of their main terms.
The renormalized spectrum of the system is obtained within the solution of dispersion equation in the vicinity of the main state where the high- and low-energy complexes of bound states are observed. The properties of the spectrum are analyzed depending on the coupling constant and the temperature.


\pacs 71.38.-k, 63.20.kd, 63.20.dk, 72.10.Di
\end{abstract}

\section{Introduction}

\looseness=-1 Obviously, there are many reasons that despite of the long period of its successful development, the theory of the interaction of quasi-particles (electrons, excitons and so on) with quantized fields (phonons, photons) remains in the center of physicists attention.  One of them is that the new phenomena permanently discovered in physics (superfluidity, superconductivity, high-temperature superconductivity and so on) in the classic 3d  and 2d structures \cite{Dav76, Abr12, Mis09} and, further \cite{Gio09, tka15} in low-dimensional heterostructures (in particular, nano-structures) demanded a deep understanding of the specific interaction between quasi-particles proper or with different fields and urged the development of mathematical methods in theoretical physics.

\looseness=-1 A fundamental progress in the theory of interaction of quasi-particles-quantized fields in condensed materials has been achieved late in the twentieth century when the methods of quantum field theory were developed \cite{Fro50, Fey62}, in particular, the Feynman diagram technique in the method of thermodynamic Green's functions \cite{Abr12, Tka03, Sta13}. However, this universal and powerful mathematical method is not devoid of the difficulties at the stage of its direct application to the specific problems where the perturbation method or the variational one is not applicable. For instance, the difficulties associated with the peculiar problem of sign \cite{Mis05} appear when the excited hybrid states of the system \cite{Lev73} are studied and it is necessary to take into account the multi-phonon processes. In any representation, except the Matsubara one, the expanded Green's function in the representation of interaction contains complex terms with the sign that changes in general case. Hence, the mathematical structure of mass operator (MO) ranges becomes complicated. Therefore, in order to reliably define  the spectrum of a quasi-particle interacting with phonons in a wide range of energies, it is not enough to account for the finite number of diagrams, but it is necessary to perform a partial summing of an infinite number of diagrams, if not all of them, then at least the main ones.

The long period of a lack of reliable and accurate data on the main and excited states in macroscopic structures has stimulated a search for new approaches to the mathematical methods of their research. This, in particular, contributed to the creation of diagrammatic Monte Carlo method, which together with the method of stochastic optimization made it possible to calculate the Green's functions of quasi-particles in different models \cite{Mis05, Pro98, Prok98, Mis00, Ber06, Ber10, Mar10, Ebr12, Fil12, Gou16, Gou17} almost exactly and to get reliable data where they were incomplete or very controversial \cite{Mis05}, like in the concept of the relaxed excited states. This method does not require any explicit ``circumcision'' of the orders of the ranges of Green's functions \cite{Mis05} and satisfies the central limit theorem, since, using a sufficient reserve of memory of modern computers, one can always achieve an insignificantly small effect of the system fluctuations and obtain an almost exact result.

Modern computers significantly expanded the scope of applying the Feynman diagram technique in the method of Green's functions for the problems of quasi-particles-phonons interaction, especially in the cases where it is necessary to consider the multi-phonon processes \cite{Mis09, Lev73, Mis05}. Recently, using a modified  Feynman-Pines diagram technique, computer calculations of the leading classes of MO diagrams were performed \cite{Tka15}. Herein, a physically correct picture was established, which was missing earlier, for the ground and excited states of Frohlich polaron in the regime of a weak electron-phonon coupling at $T=0$~K.

A renormalized spectrum of a localized quasiparticle weakly \, interacting \, with \, phonons \,at $T\neq0$~K was obtained in \cite{Tka16} using the Feynman-Pines diagram technique in the general model with the Frohlich Hamiltonian. It was shown that the new complexes of bound states existed in the system besides the ground and bound states which were known earlier \cite{Dav76, tka15} from the simplified model with the additional condition for the operators of quasi-particles. The energy spectrum was calculated using the MO, which contained all diagrams till the third order. Of course, this was not enough to identify the leading diagrams for their partial summing. However, it was impossible to hope for the analytical calculation of high-order diagrams without a computer, since their number in the $n$-th order was proportional to $n\cdot2^{n}(2n-1)!!$.

In this paper, we use a new  computer program for  analytical calculation of MO in such a high order over the coupling constant, which is limited by the computer resource only. Although we take into account all diagrams till the tens order, using a personal computer, this is enough for a partial summing of the main MO diagrams. As a result, we confirm the existence of previously known and new states of a quasiparticle interacting with phonons and analyze the temperature dependence of the spectrum.

\section{Hamiltonian of the system. Multi-phonon processes and structure of complete MO at $T\neq0$~K}

Localized dispersionless quasiparticles (excitons, impurities and so on) interacting with dispersionless polarization phonons are described by Frohlich Hamiltonian, like in \cite{Tka16}
\begin{equation} \label{GrindEQ__1_}
H=E\sum _{\vec{k}}{\widehat{a}}_{\vec{k}}^{+} {\widehat{a}}_{\vec{k}}+{\Omega}\sum _{\vec{q}}\left(\widehat{b}_{\vec{q}}^{+} {\widehat{b}}_{\vec{q}} +\frac{1}{2}\right)+\sum _{\vec{k}\vec{q}}\varphi(q)\widehat{a}_{\vec{k}+\vec{q}}^{+}{\widehat{a}}_{\vec{k}}\left({\widehat{b}}_{\vec{q}} +{\widehat{b}}_{-\vec{q}}^{+}\right),
\end{equation}
where $E$ and  $\Omega$ are the energies of quasi-particles and phonons, respectively, which can be arbitrary in the typical ranges ($E\approx 100\div 1000$~meV, $\Omega \approx 20\div 100$~meV) for the solid states and nano-structures. It turns out that the form of $\varphi(q)$ binding function is irrelevant because the energy of quasi-particle-phonon interaction is uniquely characterized by a constant $\alpha $ independent of $\vec{q}$ , also arbitrary in the natural range, since the problem becomes zero-dimensional. The operators of second quantization of quasi-particles ($\widehat{a}_{\vec{k}},\widehat{a}_{\vec{k}}^{+}$) and phonons ($\widehat{b}_{\vec{q}}, \widehat{b}_{\vec{q}}^{+} $) satisfy the Bose commutative relationships.

\begin{figure}[!b]
\centerline{\includegraphics[width=0.8\textwidth]{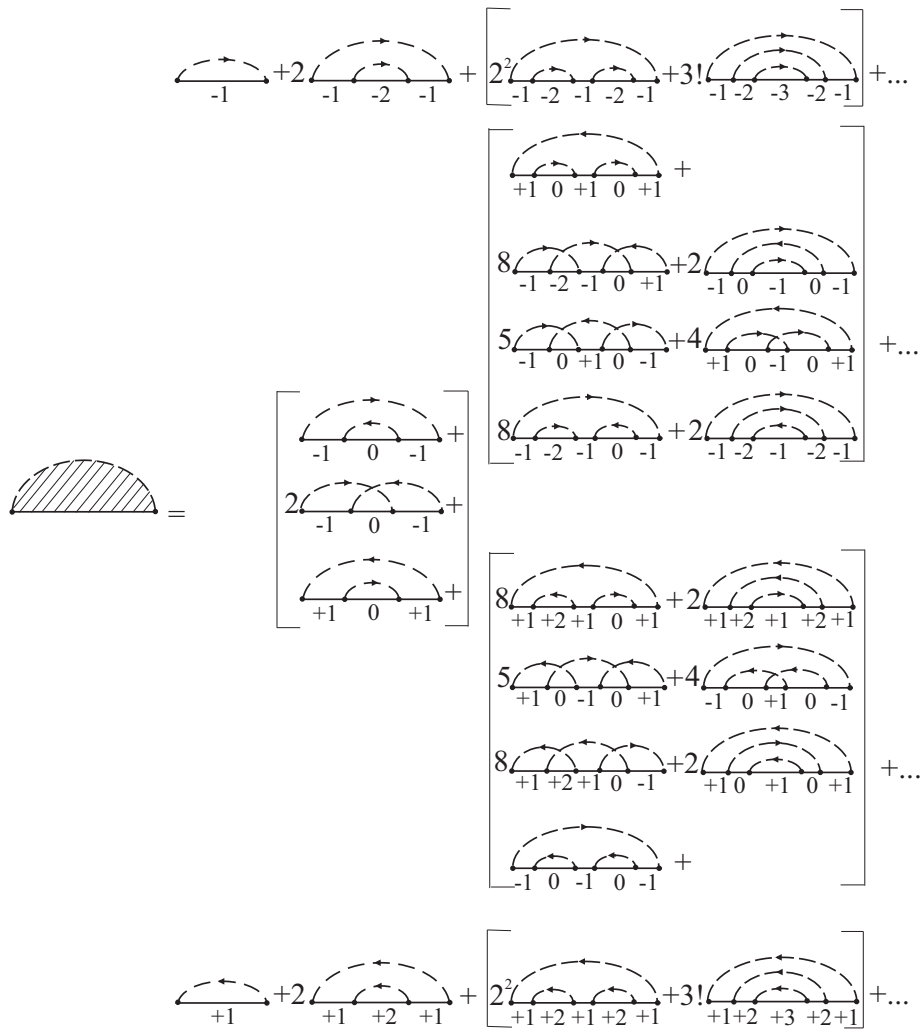}}
\caption{\label{fig1} Diagrammatic representation of MO terms in the first three orders of $\alpha$ powers.}
\end{figure}

The renormalized spectrum of quasi-particles interacting with phonons at an arbitrary temperature~($T$), like in \cite{Tka16}, is obtained from the poles of Fourier image of quasi-particle Green's function $G(\vec{k},\omega')$, which through the Dyson equation
\begin{equation} \label{GrindEQ__2_}
G(\vec{k},\omega^{\prime})=\{\omega^{\prime}-E(\vec{k})-M(\vec{k},\omega^{\prime})\}^{-1}, \qquad(\hbar =1,\ \omega^{\prime}=\omega +\ri\eta )
\end{equation}
is related with MO $M(\vec{k},\omega')$. It is calculated using the Feynman-Pines diagram technique when the concentration of localized quasi-particles is small. Introducing the dimensionless variables and values
\begin{equation} \label{GrindEQ__3_}
\xi=\frac{\omega-E}{\Omega}\,,\qquad\alpha=\sum_{\vec{q}}\frac{|\varphi(q)|^{2}}{\Omega^{2}}\,,\qquad{\Large\textsl{m}}=\frac{M}{\Omega}\,,\qquad g=\Omega G,
\end{equation}
the Dyson equation \eqref{GrindEQ__2_} is rewritten as
\begin{equation} \label{GrindEQ__4_}
g(\xi)=\{\xi-{\Large\textsl{m}}(\xi)\}^{-1}.
\end{equation}
As far as the problem becomes ``zero-dimensional'', the complete MO ${\Large\textsl{m}}(\xi)$ is defined by the infinite sum of indexed diagrams (figure~\ref{fig1}). The simple rules for these diagrams presented in  \cite{Tka16}, allow their definite programming and numerical calculation of all topologically unequal diagrams and their analytical expressions till such a high order over the powers of a coupling constant ($\alpha $), which is limited by a computer resource only (appendix~\ref{App1}). We should note that in this paper all diagrams and their analytical expressions for the MO terms till the tenth order are obtained. Although in figure~\ref{fig1}, all first terms of MO, including the third order, are presented for  illustration and for further understanding.

The rules of conformity between indexed diagrams and analytical expressions are simple:
\begin{equation} \label{GrindEQ__5_}
\raisebox{0pt}{\hfil\includegraphics[width=0.9\textwidth]{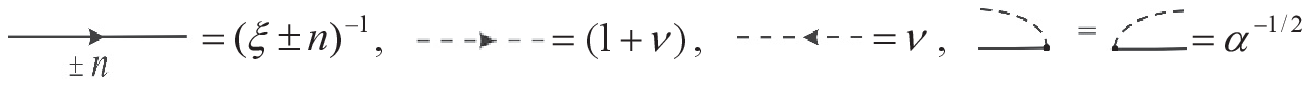}}\raisebox{12pt}{$\hspace{-2mm},$}
\end{equation}
where number $n$ is the sum of phonon lines with the arrows directed to the right (with the sign ``$-$'') and with the arrows directed to the left (with the sign ``$+$'') over the quasi-particle line (figure~\ref{fig1}), $\nu =(\re^{\Omega /kT} -1)^{-1} $ is an average phonon occupation number.
Thus, the analytical expression for the arbitrary diagram with indices is written as a product of contributions of all its lines and tops. For example, the analytical expressions are as follows:
\begin{equation} \label{GrindEQ__6_}
\raisebox{0pt}{\hfil\includegraphics[width=0.8\textwidth]{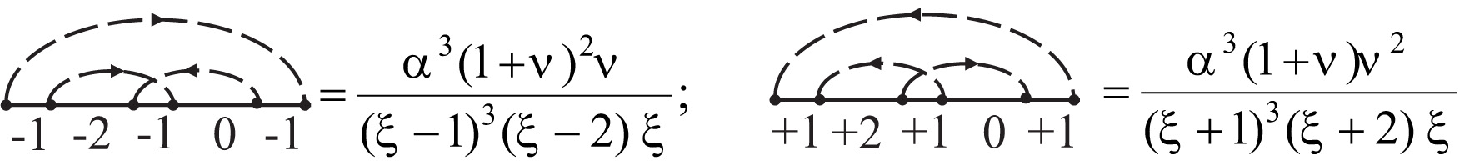}}\raisebox{16pt}{\hspace{1mm}.}
\end{equation}

A detailed analysis of the structure of complete MO, figure~\ref{fig1}, till  high orders over the powers of $\alpha $, $\nu $, (1+$\nu $) shows that it can be written in an analytical form which contains three types of terms. The terms with the symbol ``$>$'': ${\Large\textsl{m}}_{k,\, 2k+n}^{>} (\xi )\,\sim \alpha^{2k+n}\nu ^{k} (1+\nu )^{k+n}$ $(k=0,1,\dots,$ $n=1,2,\dots)$ describe the processes of quasi-particle scattering accompanied by the prevailing creation of phonons, the terms with the symbol ``$<$'': ${\Large\textsl{m}}_{k,\, 2k+n}^{<} (\xi )\, \sim\alpha ^{2k+n} \nu ^{k+n} (1+\nu )^{k} $ describe the processes of quasi-particle scattering accompanied by the prevailing annihilation of phonons and ${\Large\textsl{m}}^{(\text e)}_{2k} (\xi )\sim\alpha ^{2k} [\nu (1+\nu )]^{k} $. Hence, the infinite ranges of complete MO terms can be written as
\begin{equation} \label{GrindEQ__7_}
{\Large\textsl{m}}(\xi )=\left\{\begin{array}{ccccccc} 
{{\Large\textsl{m}}_{\text u}^{>} (\xi )=} & {{\Large\textsl{m}}_{0,1}^{>} (\xi )+}  & {{\Large\textsl{m}}_{0,2}^{>} (\xi )+} & {{\Large\textsl{m}}_{0,3}^{>} (\xi )+} & {{\Large\textsl{m}}_{0,4}^{>} (\xi )+} & {{\Large\textsl{m}}_{0,5}^{>} (\xi )} & {+\ldots} \\
 {{\Large\textsl{m}}_{1}^{>} (\xi )=}  & {} & {} & {{\Large\textsl{m}}_{1,3}^{>} (\xi )+} & {{\Large\textsl{m}}_{1,4}^{>} (\xi )+} & {{\Large\textsl{m}}_{1,5}^{>} (\xi )} & {+\ldots} \\ 
 {{\Large\textsl{m}}_{2}^{>} (\xi )=} & {} & {} & {} & {} & {{\Large\textsl{m}}_{2,5}^{>} (\xi )} & {+\ldots} \\ 
 {\ldots}  & {} & {} & {} & {} & {} & {+\ldots} \\ 
 {{\Large\textsl{m}}_{k}^{>} (\xi )=}  & {} & {} & {} & {} & {} & {+\ldots} \\ 
 {\ldots}  & {} & {} & {} & {} & {} & {+\ldots} \\ 
 {{\Large\textsl{m}}^{(\text e)} (\xi )=} & {} & {{\Large\textsl{m}}^{(\text e)}_{2} (\xi )+} & {} & {{\Large\textsl{m}}^{(\text e)}_{4} (\xi )}  & {} & {+\ldots} \\ 
 {\ldots}  & {} & {} & {} & {} & {} & {+\ldots} \\ 
 {{\Large\textsl{m}}_{k}^{<} (\xi )=} & {} & {} & {} & {} & {} & {+\ldots} \\ 
 {\ldots}  & {} & {} & {} & {} & {} & {+\ldots} \\ {{\Large\textsl{m}}_{2}^{<} (\xi )=} & {} & {} & {} & {\qquad \quad \,\,\, } & {{\Large\textsl{m}}_{2,5}^{<} (\xi )} & {+\ldots} \\ 
 {{\Large\textsl{m}}_{1}^{<} (\xi )=} & {} & {} & {{\Large\textsl{m}}_{1,3}^{<} (\xi )+} & {{\Large\textsl{m}}_{1,4}^{<} (\xi )+} & {{\Large\textsl{m}}_{1,5}^{<} (\xi )} & {+\ldots} \\ 
 {{\Large\textsl{m}}_{\text u}^{<} (\xi )=} & {{\Large\textsl{m}}_{0,1}^{<} (\xi )+} & {{\Large\textsl{m}}_{0,2}^{<} (\xi )+} & {{\Large\textsl{m}}_{0,3}^{<} (\xi )+} & {{\Large\textsl{m}}_{0,4}^{<} (\xi )+} & {{\Large\textsl{m}}_{0,5}^{<} (\xi )} & {+\ldots} \end{array}\right..
\end{equation}

In this formula, the upper ${\Large\textsl{m}}_{\text u}^{>} (\xi )$ and the lower ${\Large\textsl{m}}_{\text u}^{<} (\xi )$ terms of the complete MO are defined by the respective infinite ranges of the terms ${\Large\textsl{m}}_{0,n}^{\gtrless} (\xi )$, to which the diagrams without crossing phonon lines with arrays directed only to the right ($>$) and only to the left ($<$) correspond. These diagrams describe the unmixed (u) processes of quasi-particles scattering accompanied either only by the creation ($>$) or by annihilation ($<$) of phonons, respectively. After the partial summing, an exact analytical representation for these terms in the form of infinite chain fractions was obtained \cite{Tka16}.
\begin{equation} \label{GrindEQ__8_}
{\Large\textsl{m}}_{\text u} (\xi )={\Large\textsl{m}}_{\text u}^{>} (\xi )+{\Large\textsl{m}}_{\text u}^{<} (\xi ),
\end{equation}
where
\begin{equation} \label{GrindEQ__9_}
{\rm {\Large\textsl{m}}}_{1}^{>} (\xi )=\frac{\alpha (1+\nu )}{\xi-1-\displaystyle\frac{2 \alpha (1+\nu )}{\xi -2-\ldots-\displaystyle\frac{n \alpha  (1+\nu )}{\xi -n-\ldots} }}\,,\quad{\Large\textsl{m}}_{1}^{<} (\xi )=\frac{\alpha \nu }{\xi+1-\displaystyle\frac{2 \alpha \nu }{\xi +2-\ldots-\displaystyle\frac{n \alpha \nu }{\xi +n-\ldots} } }\,.
\end{equation}
The rest terms of MO \eqref{GrindEQ__7_}
\begin{equation} \label{GrindEQ__10_}
{\Large\textsl{m}}_{\text m} (\xi )={\Large\textsl{m}}^{(\text e)} (\xi )+{\Large\textsl{m}}^{>} (\xi )+{\Large\textsl{m}}^{<} (\xi )
\end{equation}
describe the mixed (m) processes of quasi-particle scattering accompanied by an equal (e) number of the created and annihilated phonons
\begin{equation} \label{GrindEQ__11_}
{\Large\textsl{m}}^{(\text e)} (\xi )=\sum _{k=1}^{\infty }{\Large\textsl{m}}^{(\text e)}_{2k} (\xi ) =\frac{\sum\limits_{k=1}^{\infty }\alpha ^{2k} [\nu (1+\nu )]^{k}  \varphi_{\text{(e)}2k} (\xi )}{\prod\limits_{l=1}^{k}(\xi ^{2} -l^{2} ) ^{2k+1-l}}\,,
\end{equation}
by a prevailing number of the created phonons rather than annihilated ones
\begin{align} \label{GrindEQ__12_}
{\Large\textsl{m}}^{>} (\xi )&=\sum _{k=1}^{\infty }{\Large\textsl{m}}_{k}^{>} (\xi ) =\sum _{k=1}^{\infty }\sum _{n=1}^{\infty }{\Large\textsl{m}}_{k,2k+n}^{>} (\xi )
\nonumber\\
&=\sum _{k,n=1}^{\infty }\frac{\alpha ^{2k+n} (1+\nu )^{k} \nu ^{k+n} \varphi _{k,2k+n}^{>} (\xi )}{\prod\limits_{l_{1} =1}^{k}(\xi ^{2} -l_{1}^{2} ) ^{2k+n+1-l_{1} } \prod\limits_{l_{2} =1}^{n}[\xi -(k+l_{2} )] ^{n+k+1-l_{2} } }\,,
\end{align}
and by a prevailing number of annihilated phonons rather than the created ones
\begin{align} \label{GrindEQ__13_}
{\Large\textsl{m}}^{<} (\xi )&=\sum _{k=1}^{\infty }{\Large\textsl{m}}_{k}^{<} (\xi ) =\sum _{l=1}^{\infty }\sum _{n=1}^{\infty }{\Large\textsl{m}}_{l,2k+n}^{<} (\xi )
\nonumber\\
&=\sum _{l,n=1}^{\infty }\frac{\alpha ^{2k+n} (1+\nu )^{k} \nu ^{l+n} \varphi _{k,2k+1}^{<} (\xi )}{\prod\limits_{l_{1} =1}^{k}(\xi ^{2} -l_{1}^{2} ) ^{2k+n+1-l_{1} } \prod\limits_{l_{2} =1}^{n}[\xi -(k+l_{2})] ^{n+k+1-l_{2} } }\,.
\end{align}
The functions $\varphi _{\text{(e)}2k} (\xi )=-\varphi _{\text{(e)}2k} (-\xi )$ and $\varphi _{k,2l+n}^{>} (\xi )=-\varphi _{k,2l+n}^{<} (-\xi )$ are  polynomials,  diagrammatic and analytical forms of which are numerically calculated within the computer using the diagram technique. Till the sixth order over the coupling constant ($\alpha $), the expressions for these functions are presented in  appendix~\ref{App2}, though in this paper we took into account all diagrams till the tenth order.

Finally, the complete MO \eqref{GrindEQ__7_} now has the form
\begin{equation} \label{GrindEQ__14_}
{\Large\textsl{m}}(\xi )={\Large\textsl{m}}_{\text u}^{} (\xi )+{\Large\textsl{m}}_{\text m}^{} (\xi ),
\end{equation}
where
\begin{equation} \label{GrindEQ__15_}
{\Large\textsl{m}}_{\text u}^{>} (\xi ,1+\nu )=-{\Large\textsl{m}}_{\text u}^{<} (\xi ,\nu ),\qquad{\Large\textsl{m}}_{}^{>} (\xi ;1+\nu ,\nu )=-{\Large\textsl{m}}_{}^{<} (-\xi ;\nu ,1+\nu ).
\end{equation}

Contrary to the unmixed terms \eqref{GrindEQ__9_}, the partial summing in the mixed ones~\eqref{GrindEQ__11_}--\eqref{GrindEQ__13_}
is, in general, impossible. However, using the newly revealed analytical structure, this becomes possible for the arbitrary energies, highlighting in these MO terms the pole and non-pole factors and expanding them in Taylor series.

\section{Partial summing in mixed MO using the non-pole factors expanded in Taylor series}

For a system having a weak quasi-particle-phonon interaction ($\alpha<1$) at the temperature when $\nu\ll1$, we observe the range of energies ($-2<\xi <2$) where, according to physical considerations, one should expect the renormalized energies of the main state and its nearest satellite bound-with-phonon states. In order to obtain them, both in high- and low-energy regions, we perform a partial summing in mixed MO~\eqref{GrindEQ__11_}--\eqref{GrindEQ__13_}.

First of all, we observe ${\Large\textsl{m}}^{>} (\xi )$ in the range $0<\xi<2$, selecting pole and non-pole factors in it. Formula~\eqref{GrindEQ__11_} proves that MO terms MO ${\Large\textsl{m}}_{k,\, 2k+n}^{>} (\xi )$ in this energy range can be submitted as a product
\begin{equation} \label{GrindEQ__16_}
{\Large\textsl{m}}_{k,\, 2k+n}^{>} (0<\xi <2)=\frac{\alpha ^{2k+n} \nu ^{k} (1+\nu )^{k+n} }{(\xi -1)^{2k+n} } \, f_{k,2k+n}^{>} (\xi ),\qquad k,n=1,2,3,\ldots,
\end{equation}
where one can see a multiplier independent of $\xi$, pole multiplier and non-pole fractional-rational functions
\begin{equation} \label{GrindEQ__17_}
f_{1,2+n}^{>} (\xi )=\varphi _{1,2+n}^{>} (\xi )\bigg\{(\xi ^{2} +1)^{2+n} \prod _{l=1}^{n}[\xi -(l+1)]^{2+n-l}  \bigg\}^{-1},
\end{equation}
\begin{equation} \label{GrindEQ__18_}
f_{k,2k+n}^{>} (\xi )=\frac{\varphi _{k,2k+n}^{>} (\xi )}{(\xi +1)^{2k+n} \prod\limits_{l_{1} =1}^{k-1}[\xi ^{2} -(l_{1} +1)^{2} ]^{2k+n-l_{1} } \prod\limits_{l_{2} =1}^{k}[\xi -(k+l_{2} )]^{k+n+1-l_{2} } }\,, \ \  \left\{\begin{array}{l} {k=2,3,\ldots} \\ {n=1,2,3,\ldots} \end{array},\right.
\end{equation}
which in the vicinity of $\xi=1$ are expanded into the Taylor series
\begin{equation} \label{GrindEQ__19_}
f_{k,2k+n}^{>} (\xi )=\sum _{s=0}^{N_{k,n} }F_{k,2k+n}^{>(s)}(1)(\xi -1)^{s},\qquad\left(F_{k,2n}^{>(s)}(1)=\left. \frac{\partial ^{s} f_{k,2k+n}^{>} (\xi )}{s!\partial \xi ^{s} } \right|_{\xi =1} \right),\,\,\,(k,n=1,2,\ldots),
\end{equation}
where $F_{k,2n}^{>(s)} $ are the known numerical coefficients.

Substituting \eqref{GrindEQ__17_}, \eqref{GrindEQ__18_} in \eqref{GrindEQ__14_} we see that now MO ${\Large\textsl{m}}_{k,\, 2k+n}^{>}$ $(0<\xi <2)$ can be written as a sum of pole ${\Large\textsl{m}}_{k,\, 2k+n}^{\text{(p)}>} (\xi )$ and non-pole ${\Large\textsl{m}}_{k,\, 2k+n}^{\text{(np)}>} (\xi )$ terms
\begin{align} \label{GrindEQ__20_}
&{\Large\textsl{m}}_{k,\, 2k+n}^{>} (0<\xi <2)={\Large\textsl{m}}_{k,\, 2k+n}^{\text{(p)}>} (\xi )+{\Large\textsl{m}}_{k,\, 2k+n}^{\text{(np)}>} (\xi )
\nonumber\\
&=\alpha ^{2k+n} \nu ^{k} (1+\nu )^{k+n} \, \left(\sum _{s=0}^{2k+n-1}+\sum _{s=2k+n}^{\infty }\right)F_{k,2k+n}^{>(s)} (1)(\xi -1)^{s-(2k+n)}.
\end{align}
Similar analytics gives
\begin{align} \label{GrindEQ__21_}
&{\Large\textsl{m}}_{k,\, 2k+n}^{<} (0<\xi <2)={\Large\textsl{m}}_{l,\, 2k+n}^{\text{(p)}<} (\xi )+{\Large\textsl{m}}_{l,\, 2k+n}^{\text{(np)}<} (\xi )
\nonumber\\
&=\alpha ^{2k+n} \nu ^{k+n} (1+\nu )^{k} \, \left(\sum _{s=0}^{2k+n-1}+\sum _{s=2k+n}^{\infty } \right)F_{l,2k+n}^{<(s)} (1)(\xi -1)^{s-(2k+n)} ,
\end{align}
\begin{equation} \label{GrindEQ__22_}
{\Large\textsl{m}}_{2k} (0<\xi <2)={\Large\textsl{m}}_{2k}^{\text{(p)}} (\xi )+{\Large\textsl{m}}_{2k}^{\text{(np)}} (\xi )=\alpha ^{2k} [\nu (1+\nu )]^{k}\left(\sum _{s=0}^{2k-1}+\sum _{s=2k}^{\infty } \right)F_{2k}^{(s)} (1)(\xi -1)^{s-2k},
\end{equation}
where
\begin{equation} \label{GrindEQ__23_}
F_{k,2k+n}^{>(s)} (1)=\left. \frac{\partial ^{s} f_{k,2k+n}^{>} (\xi )}{s!\partial \xi ^{s} } \right|_{\xi =1},\qquad F_{2k}^{(s)} (1)=\left. \frac{\partial ^{s} f_{2k}^{} (\xi )}{s!\partial \xi ^{s} } \right|_{\xi =1}.
\end{equation}

Substituting formulae~\eqref{GrindEQ__20_}--\eqref{GrindEQ__22_} in \eqref{GrindEQ__11_}--\eqref{GrindEQ__13_} we select pole and non-pole terms in the mixed MO ${\Large\textsl{m}}^{(\text e)} (0<\xi <2)$  and ${\Large\textsl{m}}^{>} (0<\xi <2)$, ${\Large\textsl{m}}^{<} (0<\xi <2)$  in the range ($0<\xi<2$). Taking into account formula~\eqref{GrindEQ__15_}, they are obtained in the range ($-2<\xi<0$). Thus, according to \eqref{GrindEQ__10_}, a complete mixed MO ${\Large\textsl{m}}_{\text m} (\xi )$ is written as the sum of pole ${\Large\textsl{m}}^{\text{(p)}} (\xi )$ and non-pole ${\Large\textsl{m}}^{\text{(np)}} (\xi )$ term
\begin{align} \label{GrindEQ__24_}
&{\Large\textsl{m}}_{\text m} (-2<\xi <2)=\sum _{k=1}^{\infty }{\Large\textsl{m}}_{k}  (-2<\xi <2)=\sum _{k=1}^{\infty }\left[{\Large\textsl{m}}_{k}^{\text{(p)}} (-2<\xi <2)+{\Large\textsl{m}}_{k}^{\text{(np)}} (-2<\xi <2)\right]
\nonumber\\
&=\sum _{k=1}^{\infty }\left\{\left[{\Large\textsl{m}}_{2k}^{\text{(p)}}(\xi )+{\Large\textsl{m}}_{k}^{\text{(p)}>} (\xi )+{\Large\textsl{m}}_{k}^{\text{(p)}<} (\xi )\right]+\left[{\Large\textsl{m}}_{2k}^{\text{(np)}} (\xi )+{\Large\textsl{m}}_{k}^{\text{(np)}>} (\xi )+{\Large\textsl{m}}_{k}^{\text{(np)}<} (\xi )\right]\right\}.
\end{align}

The expression for ${\Large\textsl{m}}_{\text m}(\xi )$ obtained in the form of separated pole and non-pole terms gives an opportunity to perform a partial summing of their infinite ranges. It is clear that since the maximal order ($k_{\text{max}}$) of functions $f_{k,2k+n} $ is limited by the maximal order of the accounted diagrams over $\alpha $, then in the complete ${\Large\textsl{m}}_{\text m} (\xi )$ one should take into account not more than $k_{\text{max}}$ (in our case $k_{\text{max}}=10$) mixed renormalized MOs [${\Large\textsl{m}}_{k}^{\text{(p)}}(\xi ), {\Large\textsl{m}}_{k}^{\text{(np)}} (\xi )$]. The sums over $k$ quickly converge because $k$-th terms of this MO contain  small multipliers $\alpha^{k}$ and $\nu^{k}$. Avoiding  sophisticated analytics, considering the symmetric relationship \eqref{GrindEQ__15_} and a similar structure of ${\Large\textsl{m}}_{k} $ at all $k$ values, we demonstrate the method of partial summing by the example of MO ${\Large\textsl{m}}_{1} (\xi )$ in the range ($0<\xi<2$).

In table~\ref{tbl-smp1} we present  explicit analytical expressions for ${\Large\textsl{m}}_{1} (\xi )$ terms. Solid lines separate pole and non-pole terms. It is clear that the pole terms of ${\Large\textsl{m}}_{1}^{\text{(p)}} (\xi )$ can be written as a sum of ${\Large\textsl{m}}_{2}^{\text{(p)}(0)} (\xi )$ and ${\Large\textsl{m}}_{2}^{\text{(p)}(s)} (\xi )$ terms obtained within the partial summing of the pole terms ${\Large\textsl{m}}_{1,2+n}^{\text{(p)}(s)>} (\xi )$ and ${\Large\textsl{m}}_{1,2+n}^{\text{(p)}(s)<} (\xi )$, only
\begin{equation} \label{GrindEQ__25_}
{\Large\textsl{m}}_{1}^{\text{(p)}} (\xi )=\sum _{s=0}^{9}{\Large\textsl{m}}_{1}^{\text{(p)}(s)} (\xi ) =\sum _{s=0}^{9}\left\{{\Large\textsl{m}}_{2}^{\text{(p)}(s)} (\xi )+\sum _{n=s-1}^{\infty }\left[{\Large\textsl{m}}_{1}^{>\text{(p)}(s)} (\xi )+{\Large\textsl{m}}_{1}^{<\text{(p)}(s)} (\xi )\right] \right\} ,
\end{equation}
where
\begin{equation} \label{GrindEQ__26_}
{\Large\textsl{m}}_{2}^{\text{(p)}(s\ne 0)} (\xi )={\Large\textsl{m}}_{1,2}^{\text{(p)}(1)} (\xi )=0.
\end{equation}

\begin{table}[!t]
\vspace{-1ex}
\caption{The terms of MO ${\Large\textsl{m}}_{1}(\xi)$ in the range $(0<\xi<2$). }
\label{tbl-smp1}
\begin{center}
\begin{tabular}{|p{2.8cm}|@{}p{1cm}@{}|@{}p{1cm}@{}|@{}p{1.3cm}@{}|@{}p{1.3cm}@{}|@{}p{1.3cm}@{}|@{}p{1.8cm}@{}|@{}p{1.75cm}@{}|@{}p{1.3cm}@{}|}
\hline
\hline
\qquad\quad $s$&\quad 0&\quad 1&\quad \,\,2&\quad \,\,3&\quad \,\,4&\quad \quad 5&\quad \quad 6&\\
\hline
\hline
$m_{1,2+8}^{>(s)}(\xi)=$

$\alpha\nu[\alpha(1+\nu)]^{9}\times$
      &  $\frac{2304}{(1-\xi)^{10}}$&  $\frac{-5120}{(1-\xi)^{9}}$&  $\frac{6848}{(1-\xi)^{8}}$&  $\frac{-7056}{(1-\xi)^{7}}$&  $\frac{53260}{9(1-\xi)^{6}}$&  $\frac{-22891}{6(1-\xi)^{5}}$&  $\frac{3083107}{2700(1-\xi)^{4}}$& \strut\\ \hline
$m_{1,2+7}^{>(s)}(\xi)=$

$\alpha\nu[\alpha(1+\nu)]^{8}\times$
      &  $\frac{1024}{(1-\xi)^{9}}$&  $\frac{-2016}{(1-\xi)^{8}}$&  $\frac{2400}{(1-\xi)^{7}}$& $\frac{-2186}{(1-\xi)^{6}}$& $\frac{13949}{9(1-\xi)^{5}}$&  $\frac{-121601}{180(1-\xi)^{4}}$&  $\frac{-2522741}{8640(1-\xi)^{3}}$&\strut\\
\hline
$m_{1,2+6}^{>(s)}(\xi)=$

$\alpha\nu[\alpha(1+\nu)]^{7}\times$
      & $\frac{448}{(1-\xi)^{8}}$&  $\frac{-768}{(1-\xi)^{7}}$&  $\frac{796}{(1-\xi)^{6}}$& $\frac{-1852}{3(1-\xi)^{5}}$&  $\frac{1317}{4(1-\xi)^{4}}$&  $\frac{-4853}{2160(1-\xi)^{3}}$&  $\frac{-182879691}{259200(1-\xi)^{2}}$&\strut \\
\hline
$m_{1,2+5}^{>(s)}(\xi)=$

$\alpha\nu[\alpha(1+\nu)]^{6}\times$
      & $\frac{192}{(1-\xi)^{7}}$&  $\frac{-280}{(1-\xi)^{6}}$&  $\frac{244}{(1-\xi)^{5}}$& $\frac{-149}{(1-\xi)^{4}}$&  $\frac{320}{9(1-\xi)^{3}}$&  $\frac{107983}{1440(1-\xi)^{2}}$& $\,\,\,\frac{-99457}{576(1-\xi)}$&\strut\\
\hline \cline{8-8}
\end{tabular}
\begin{tabular}{|p{2.8cm}|@{}p{1cm}@{}|@{}p{1cm}@{}|@{}p{1.3cm}@{}|@{}p{1.3cm}@{}|@{}p{1.3cm}@{}|@{}p{1.8cm}@{}@{\setlength{\arrayrulewidth}{1pt}\vline}@{}p{1.75cm}@{}|@{}p{1.3cm}@{}|}
$m_{1,2+4}^{>(s)}(\xi)=$

$\alpha\nu[\alpha(1+\nu)]^{5}\times$
      & $\frac{80}{(1-\xi)^{6}}$&  $\frac{-96}{(1-\xi)^{5}}$&  $\frac{66}{(1-\xi)^{4}}$& $\frac{-25}{(1-\xi)^{3}}$&  $\frac{-2069}{144(1-\xi)^{2}}$&  $\quad\frac{3439}{72(1-\xi)}$\quad& \quad$\frac{-242069}{3456}$
      & $m_{1,2+4}^{\text{(np)}>}$ \strut\\
\hline \cline{7-7}
\end{tabular}
\begin{tabular}{|p{2.8cm}|@{}p{1cm}@{}|@{}p{1cm}@{}|@{}p{1.3cm}@{}|@{}p{1.3cm}@{}|@{}p{1.3cm}@{}@{\setlength{\arrayrulewidth}{1.5pt}\vline}@{}p{1.8cm}@{}|@{}p{1.75cm}@{}|@{}p{1.3cm}@{}|}
$m_{1,2+3}^{>(s)}(\xi)=$

$\alpha\nu[\alpha(1+\nu)]^{4}\times$
      & $\frac{32}{(1-\xi)^{5}}$&  $\frac{-30}{(1-\xi)^{4}}$&  $\frac{14}{(1-\xi)^{3}}$& $\frac{23}{24(1-\xi)^{2}}$&  $\frac{-205}{16(1-\xi)}$&  $\quad\frac{20953}{864}$&  $\frac{208057(1-\xi)}{5184}$& $m_{1,2+3}^{\text{(np)}>}$ \strut\\
\hline \cline{6-6}
\end{tabular}
\begin{tabular}{|p{2.8cm}|@{}p{1cm}@{}|@{}p{1cm}@{}|@{}p{1.3cm}@{}|@{}p{1.3cm}@{}@{\setlength{\arrayrulewidth}{1.5pt}\vline}@{}p{1.3cm}@{}|@{}p{1.8cm}@{}|@{}p{1.75cm}@{}|@{}p{1.3cm}@{}|}
$m_{1,2+2}^{>(s)}(\xi)=$

$\alpha\nu[\alpha(1+\nu)]^{3}\times$
      & $\frac{12}{(1-\xi)^{4}}$&  $\frac{-8}{(1-\xi)^{3}}$&  $\frac{5}{4(1-\xi)^{2}}$& $\frac{13}{4(1-\xi)}$&  $\quad-\frac{37}{8}$&  $-2(1-\xi)$&  $\frac{271(1-\xi)^{2}}{64}$& $m_{1,2+2}^{\text{(np)}>}$ \strut\\
\hline \cline{5-5}
\end{tabular}
\begin{tabular}{|p{2.8cm}|@{}p{1cm}@{}|@{}p{1cm}@{}|@{}p{1.3cm}@{}@{\setlength{\arrayrulewidth}{1.5pt}\vline}@{}p{1.3cm}@{}|@{}p{1.3cm}@{}|@{}p{1.8cm}@{}|@{}p{1.75cm}@{}|@{}p{1.3cm}@{}|}
$m_{1,2+1}^{>(s)}(\xi)=$

$\alpha\nu[\alpha(1+\nu)]^{2}\times$
      & $\frac{4}{(1-\xi)^{3}}$&  $\frac{-3}{2(1-\xi)^{2}}$&  $\frac{-3}{4(1-\xi)}$&  $\quad\frac{11}{4}$&  $\frac{9(1-\xi)}{2}$&  $\frac{219(1-\xi)^{2}}{32}$&  $\frac{545(1-\xi)^{3}}{64}$& $m_{1,2+1}^{\text{(np)}>}$ \strut\\
\hline \cline{3-4}
\end{tabular}
\begin{tabular}{|p{2.8cm}|@{}p{1cm}@{}@{\setlength{\arrayrulewidth}{1.5pt}\vline}@{}p{1cm}@{}|@{}p{1.3cm}@{}|@{}p{1.3cm}@{}|@{}p{1.3cm}@{}|@{}p{1.8cm}@{}|@{}p{1.75cm}@{}|@{}p{1.3cm}@{}|}
$m_{2}^{>(s)}(\xi)=$

$\alpha\nu[\alpha(1+\nu)]\times$
      & $\frac{1}{(1-\xi)^{2}}$&\quad0&  $\quad-\frac{1}{4}$& $\frac{1(1-\xi)}{4}$&  $\frac{-3(1-\xi)^{2}}{16}$&  $\quad\frac{1(1-\xi)^{3}}{8}$&  $\frac{-5(1-\xi)^{4}}{64}$& $m_{2}^{\text{(np)}}$ \strut\\
\hline \cline{3-4}
\end{tabular}
\begin{tabular}{|p{2.8cm}|@{}p{1cm}@{}|@{}p{1cm}@{}|@{}p{1.3cm}@{}@{\setlength{\arrayrulewidth}{1pt}\vline}@{}p{1.3cm}@{}|@{}p{1.3cm}@{}|@{}p{1.8cm}@{}|@{}p{1.75cm}@{}|@{}p{1.3cm}@{}|}
$m_{1,2+1}^{<(s)}(\xi)=$

$(\alpha\nu)^{2}[\alpha(1+\nu)]\times$
      & $\frac{1}{(\xi-1)^{3}}$&\quad0&  $\frac{-1}{12(\xi-1)}$& $\quad-\frac{7}{36}$&  $\frac{59(\xi-1)}{144}$&  $\frac{-307(1-\xi)^{2}}{648}$&  $\frac{6715(1-\xi)^{3}}{15552}$&  $m_{1,2+1}^{\text{(np)}<}$ \strut\\
\hline
$m_{1,2+2}^{<(s)}(\xi)=$

$(\alpha\nu)^{3}[\alpha(1+\nu)]\times$
      & $\frac{1}{(\xi-1)^{4}}$&\quad0&  $\frac{-1}{12(\xi-1)^{2}}$&\quad\,\,\,0&  \quad$-\frac{55}{288}$&  $\frac{1367(\xi-1)}{2592}$&  $\frac{33095(\xi-1)^{2}}{41472}$&  $m_{1,2+2}^{\text{(np)}<}$ \strut\\
\hline \cline{5-6}
\end{tabular}
\begin{tabular}{|p{2.8cm}|@{}p{1cm}@{}|@{}p{1cm}@{}|@{}p{1.3cm}@{}|@{}p{1.3cm}@{}|@{}p{1.3cm}@{}@{\setlength{\arrayrulewidth}{1.5pt}\vline}@{}p{1.8cm}@{}|@{}p{1.75cm}@{}|@{}p{1.3cm}@{}|}
$m_{1,2+3}^{<(s)}(\xi)=$

$(\alpha\nu)^{4}[\alpha(1+\nu)]\times$
      & $\frac{1}{(\xi-1)^{5}}$&\quad0&  $\frac{-1}{12(\xi-1)^{3}}$&\quad\,\,\,0&  $\frac{1}{240(\xi-1)}$&  $\,\,\,-\frac{11887}{64800}$&  $\frac{315199(1-\xi)}{518400}$&  $m_{1,2+3}^{\text{(np)}<}$ \strut\\
\hline
$m_{1,2+4}^{<(s)}(\xi)=$

$(\alpha\nu)^{5}[\alpha(1+\nu)]\times$
      & $\frac{1}{(\xi-1)^{6}}$&\quad0&  $\frac{-1}{12(\xi-1)^{4}}$&\quad\,\,\,0&  $\frac{1}{240(\xi-1)^{2}}$&\quad\quad0&  $-\frac{515369}{3110400}$&  $m_{1,2+4}^{\text{(np)}<}$\strut\\
\hline \cline{7-8}
\end{tabular}
\begin{tabular}{|p{2.8cm}|@{}p{1cm}@{}|@{}p{1cm}@{}|@{}p{1.3cm}@{}|@{}p{1.3cm}@{}|@{}p{1.3cm}@{}|@{}p{1.8cm}@{}|@{}p{1.75cm}@{}|@{}p{1.3cm}@{}|}
$m_{1,2+5}^{<(s)}(\xi)=$

$(\alpha\nu)^{6}[\alpha(1+\nu)]\times$
      & $\frac{1}{(\xi-1)^{7}}$&\quad0&  $\frac{-1}{12(\xi-1)^{5}}$&\quad\,\,\,0&  $\frac{1}{240(\xi-1)^{3}}$&\quad\quad0&  $\frac{-1}{6048(\xi-1)}$&\strut\\
\hline
$m_{1,2+6}^{<(s)}(\xi)=$

$(\alpha\nu)^{7}[\alpha(1+\nu)]\times$
      & $\frac{1}{(\xi-1)^{8}}$&\quad0&  $\frac{-1}{12(\xi-1)^{6}}$&\quad\,\,\,0&  $\frac{1}{240(\xi-1)^{4}}$&\quad\quad0&  $\frac{-1}{6048(\xi-1)^{2}}$&\strut\\
\hline
$m_{1,2+7}^{<(s)}(\xi)=$

$(\alpha\nu)^{8}[\alpha(1+\nu)]\times$
      & $\frac{1}{(\xi-1)^{9}}$&\quad0&  $\frac{-1}{12(\xi-1)^{7}}$&\quad\,\,\,0&  $\frac{1}{240(\xi-1)^{5}}$&\quad\quad0&  $\frac{-1}{6048(\xi-1)^{3}}$&\strut\\
\hline
$m_{1,2+8}^{<(s)}(\xi)=$

$(\alpha\nu)^{9}[\alpha(1+\nu)]\times$
      & $\frac{1}{(\xi-1)^{10}}$&\quad0&  $\frac{-1}{12(\xi-1)^{8}}$&\quad\,\,\,0&  $\frac{1}{240(\xi-1)^{6}}$&\quad\quad0&  $\frac{-1}{6048(\xi-1)^{4}}$&\strut\\
\hline \hline
$m_{1}^{\text{(p)}}=\sum\limits_{s=0}^{\infty} m_{1}^{\text{(p)}(s)}$
      & $m_{1}^{\text{(p)}(0)}$&  $m_{1}^{\text{(p)}(1)}$&  $\,\,m_{1}^{\text{(p)}(2)}$&  $\,\,m_{1}^{\text{(p)}(3)}$&  $\,\,m_{1}^{\text{(p)}(4)}$&  $\quad m_{1}^{\text{(p)}(5)}$& $\quad m_{1}^{\text{(p)}(6)}$&\strut\\
\hline
\hline
\end{tabular}
\end{center}
\vspace{-5mm}
\end{table}

The main idea of partial summing of the main terms of pole MO is as follows: using the $s$-th column, table~\ref{tbl-smp1}, we study the analytical regularities of ${\Large\textsl{m}}_{1,2+n}^{\gtrless} (\xi )$ terms at small values $n=1,2,\dots$  and obtain an exact analytical expression for this MO at an arbitrary $n$. Then, after the analytical summing of this infinite range, we obtain an exact analytical equation for renormalized MO ${\Large\textsl{m}}_{1}^{\text{(p)}(s)} (\xi )$, which is not limited by a finite number of accounted diagrams of the respective type. For example, at $s=0$ using  table~\ref{tbl-smp1}, we obtain

\begin{align} \label{GrindEQ__27_}
&{\Large\textsl{m}}_{1}^{\text{(p)}(0)} (0<\xi <2)=\left[{\Large\textsl{m}}_{2}^{\text{(p)}(0)} (\xi )+{\Large\textsl{m}}_{1,2}^{\text{(p)}(0)>} (\xi )\right]+{\Large\textsl{m}}_{1,2}^{\text{(p)}(0)<} (\xi )
=\frac{\alpha \nu [\alpha (1+\nu )]}{(1-\xi )^{2} } \sum _{n=0}^{\infty }n\left[\frac{2\alpha (1+\nu )}{(1-\xi )} \right] ^{n} \nonumber\\&+\frac{(\alpha \nu )^{2} [\alpha (1+\nu )]}{(\xi -1)^{3} } \sum _{n=0}^{\infty }\left[\frac{\alpha \nu }{(\xi -1)} \right] ^{n} 
=\frac{\alpha \nu [\alpha (1+\nu )]}{\left\{\xi -[1-2\alpha (1+\nu )]\right\}^{2} } +\frac{(\alpha \nu )^{2} [\alpha (1+\nu )]}{(\xi -1)^{2} [\xi -(1+\alpha \nu )]}\,.
\end{align}
The renormalized ${\Large\textsl{m}}_{1}^{\text{(p)}(s)} (\xi )$ of a higher order over $s$ are obtained in the same way. Still relatively not complicated MO at $s=1$ and $s=2$ have the form
\begin{equation} \label{GrindEQ__28_}
{\Large\textsl{m}}_{1}^{\text{(p)}(1)} (\xi )=-\frac{2\alpha \nu [\alpha (1+\nu )]^{3} }{\left\{\xi -[1-2\alpha (1+\nu )]\right\}^{3} } \left\{4+9\left[\frac{\alpha (1+\nu )}{\xi -1} \right]+6\left[\frac{\alpha (1+\nu )}{\xi -1} \right]^{2} \right\}-\frac{3}{2} \left[\frac{\alpha (1+\nu )}{\xi -1} \right]^{2},
\end{equation}

\begin{align} \label{GrindEQ__29_}
&{\Large\textsl{m}}_{1}^{\text{(p)}(2)} (\xi )=\frac{2\alpha \nu [\alpha (1+\nu )]^{4} (\xi -1)^{2} }{\left\{\xi -[1-2\alpha (1+\nu )]\right\}^{4} } \left\{-4\left[\frac{\alpha (1+\nu )}{\xi -1} \right]^{4} +4\left[\frac{\alpha (1+\nu )}{\xi -1} \right]+1\right\}-\frac{(\alpha \nu )^{2} [\alpha (1+\nu )]}{12[\xi -(1+\alpha \nu )]} 
\nonumber\\
&+\frac{\alpha \nu [\alpha (1+\nu )]^{2} }{4(\xi -1)} \left[5\frac{\alpha (1+\nu )}{\xi -1} +3\right]+\frac{4\alpha \nu [\alpha (1+\nu )]^{6} }{(\xi -1)^{4} } \left\{\left[\frac{\alpha (1+\nu )}{\xi -1} \right]^{2} -\frac{\alpha (1+\nu )}{\xi -1} +\frac{1}{2} \right\}.
\end{align}

By analogy,  similar tables can be obtained at $k=2,3,\dots,k_{\text{max}}$, from which the analytical expressions for ${\Large\textsl{m}}_{k}^{\text{(p)}(s)} (\xi )$ are further calculated. For instance, we present  table~\ref{tbl-smp2} at $k=2$. We should note that if the values of $k$ and $n$ increase and $k_{\text{max}}=10$, the number of known expressions for ${\Large\textsl{m}}_{k,\, 2k+n}^{\text{(p)}(s)} (\xi )$ linearly decreases. Thus, beginning from certain $k$ and $s$, it is impossible to obtain an exact analytical expression for an arbitrary term of the respective infinite range and, thus, to perform a partial summing. Usually, this problem is solved with the help of rather powerful computers, which can essentially increase $k_{\text{max}}$. However, the presence of two small parameters ($\alpha $ and $\nu$) in the theory, brings us to the fact that the contribution of pole MO terms rapidly decreases when $k$ and $n$ increase. Thus, as further numerical calculations prove, the above mentioned circumstances have a weak effect on the magnitudes of the renormalized energies without changing the number of complexes of bound states in the  region of energies studied.

\begin{table}[!t]
\caption{The terms of MO ${\Large\textsl{m}}_{2}(\xi)$ in the range $(0<\xi<2$). }
\label{tbl-smp2}
\begin{center}
\begin{tabular}{|p{2.9cm}|@{}p{1.1cm}@{}|@{}p{1.1cm}@{}|@{}p{1.3cm}@{}|@{}p{1.3cm}@{}|@{}p{1.4cm}@{}|@{}p{1.8cm}@{}|@{}p{1.3cm}@{}|}
\hline
\hline
\quad\qquad $s$&\quad 0&\quad 1&\quad \,\,2&\quad \,\,3&\quad \,\,4&\quad \quad 5&\quad \\
\hline
\hline

$m_{2,2+7}^{>(s)}(\xi)=$

$(\alpha\nu)^{2}[\alpha(1+\nu)]^{7}\times$
      & $\frac{-1792}{(1-\xi)^{9}}$&  $\frac{2352}{(1-\xi)^{8}}$&  $\frac{-5180}{3(1-\xi)^{7}}$& $\frac{7090}{9(1-\xi)^{5}}$&  $\frac{-27}{2(1-\xi)^{5}}$& $\frac{-5105443}{12960(1-\xi)^{4}}$& \\
\hline

$m_{2,2+6}^{>(s)}(\xi)=$

$(\alpha\nu)^{2}[\alpha(1+\nu)]^{6}\times$
      & $\frac{-672}{(1-\xi)^{8}}$&  $\frac{720}{(1-\xi)^{7}}$&  $\frac{-404}{(1-\xi)^{6}}$& $\frac{868}{9(1-\xi)^{5}}$&  $\frac{9451}{108(1-\xi)^{4}}$&  $\,\,\frac{-137081}{1080(1-\xi)^{3}}$&\\
\hline

$m_{2,2+5}^{>(s)}(\xi)=$

$(\alpha\nu)^{2}[\alpha(1+\nu)]^{5}\times$
      & $\frac{-240}{(1-\xi)^{7}}$&  $\frac{200}{(1-\xi)^{6}}$&  $\frac{-220}{3(1-\xi)^{5}}$& $\frac{-28}{3(1-\xi)^{4}}$&  $\frac{16207}{432(1-\xi)^{3}}$& $\quad\frac{-1610}{81(1-\xi)^{2}}$& \\
\hline
\end{tabular}
\begin{tabular}{|p{2.9cm}|@{}p{1.1cm}@{}|@{}p{1.1cm}@{}|@{}p{1.3cm}@{}|@{}p{1.3cm}@{}|@{}p{1.4cm}@{}|@{}p{1.8cm}@{}|@{}p{1.3cm}@{}|}
$m_{2,2+4}^{>(s)}(\xi)=$

$(\alpha\nu)^{2}[\alpha(1+\nu)]^{4}\times$
      & $\frac{-80}{(1-\xi)^{6}}$&  $\frac{48}{(1-\xi)^{5}}$&  $\frac{-19}{3(1-\xi)^{4}}$& $\frac{-173}{18(1-\xi)^{3}}$&  $\frac{70}{9(1-\xi)^{2}}$&   $\quad\frac{3641}{648(1-\xi)}$& $\,\,m_{2,2+4}^{\text{(np)}>}$\\
\hline \cline{7-7}
\end{tabular}
\begin{tabular}{|p{2.9cm}|@{}p{1.1cm}@{}|@{}p{1.1cm}@{}|@{}p{1.3cm}@{}|@{}p{1.3cm}@{}|@{}p{1.4cm}@{}@{\setlength{\arrayrulewidth}{1.5pt}\vline}@{}p{1.8cm}@{}|@{}p{1.3cm}@{}|}
$m_{2,2+3}^{>(s)}(\xi)=$

$(\alpha\nu)^{2}[\alpha(1+\nu)]^{3}\times$
      & $\frac{-24}{(1-\xi)^{5}}$&  $\frac{9}{(1-\xi)^{4}}$&  $\frac{7}{4(1-\xi)^{3}}$&  $\frac{-47}{18(1-\xi)^{2}}$&  $\frac{-161}{216(1-\xi)}$&  \quad \, $\frac{3853}{1296}$&  $\,\, m_{2,2+3}^{\text{(np)}>}$\\
\hline \cline{5-6}
\end{tabular}
\begin{tabular}{|p{2.9cm}|@{}p{1.1cm}@{}|@{}p{1.1cm}@{}|@{}p{1.3cm}@{}@{\setlength{\arrayrulewidth}{1.5pt}\vline}@{}p{1.3cm}@{}|@{}p{1.4cm}@{}|@{}p{1.8cm}@{}|@{}p{1.3cm}@{}|}
$m_{4}^{>(s)}(\xi)=$

$(\alpha\nu)^{2}[\alpha(1+\nu)]^{2}\times$
      & $\frac{-6}{(1-\xi)^{4}}$& $\frac{1}{(1-\xi)^{3}}$&  $\frac{5}{6(1-\xi)^{2}}$& \quad \,\,0&$\,\,\,\frac{-535}{216}$&  $\frac{-10123(1-\xi)}{1296}$& $\quad m_{4}^{\text{(np)}}$\\
\hline \cline{5-6}
\end{tabular}
\begin{tabular}{|p{2.9cm}|@{}p{1.1cm}@{}|@{}p{1.1cm}@{}|@{}p{1.3cm}@{}|@{}p{1.3cm}@{}|@{}p{1.4cm}@{}@{\setlength{\arrayrulewidth}{1.5pt}\vline}@{}p{1.8cm}@{}|@{}p{1.3cm}@{}|}
$m_{2,2+3}^{<(s)}(\xi)=$

$(\alpha\nu)^{3}[\alpha(1+\nu)]^{2}\times$
      & $\frac{-8}{(\xi-1)^{5}}$&  $\frac{-1}{2(\xi-1)^{4}}$& $\frac{11}{12(\xi-1)^{3}}$&  $\frac{-1}{24(\xi-1)^{2}}$&  $\frac{-5}{288(\xi-1)}$&  \quad\, $\frac{15007}{5184}$& $\,\,m_{2,2+3}^{\text{(np)}<}$\\
\hline \cline{7-7}
\end{tabular}
\begin{tabular}{|p{2.9cm}|@{}p{1.1cm}@{}|@{}p{1.1cm}@{}|@{}p{1.3cm}@{}|@{}p{1.3cm}@{}|@{}p{1.4cm}@{}|@{}p{1.8cm}@{}|@{}p{1.3cm}@{}|}
$m_{2,2+4}^{<(s)}(\xi)=$

$(\alpha\nu)^{4}[\alpha(1+\nu)]^{2}\times$
      & $\frac{-10}{(\xi-1)^{6}}$&  \quad\,\,\,0& $\frac{12}{12(\xi-1)^{4}}$&  $\frac{-2}{24(\xi-1)^{3}}$& $\frac{-5}{288(\xi-1)^{2}}$&  $\,\,\frac{13}{2400(\xi-1)}$& $\,\,m_{2,2+4}^{\text{(np)}<}$\\
\hline
\end{tabular}
\begin{tabular}{|p{2.9cm}|@{}p{1.1cm}@{}|@{}p{1.1cm}@{}|@{}p{1.3cm}@{}|@{}p{1.3cm}@{}|@{}p{1.4cm}@{}|@{}p{1.8cm}@{}|@{}p{1.3cm}@{}|}
$m_{2,2+5}^{<(s)}(\xi)=$

$(\alpha\nu)^{5}[\alpha(1+\nu)]^{2}\times$
      & $\frac{-12}{(\xi-1)^{7}}$& $\frac{1}{2(\xi-1)^{6}}$&  $\frac{13}{12(\xi-1)^{5}}$& $\frac{-3}{24(\xi-1)^{4}}$&  $\frac{-5}{288(\xi-1)^{3}}$& $\,\,\frac{18}{2400(\xi-1)^{2}}$&  \\
\hline
$m_{2,2+6}^{<(s)}(\xi)=$

$(\alpha\nu)^{6}[\alpha(1+\nu)]^{2}\times$
      & $\frac{-14}{(\xi-1)^{8}}$& $\frac{1}{(\xi-1)^{7}}$&  $\frac{14}{12(\xi-1)^{6}}$& $\frac{-4}{24(\xi-1)^{5}}$&   $\frac{-5}{288(\xi-1)^{4}}$&  $\,\,\frac{23}{2400(\xi-1)^{3}}$&  \\
\hline
$m_{2,2+7}^{<(s)}(\xi)=$

$(\alpha\nu)^{7}[\alpha(1+\nu)]^{2}\times$
      & $\frac{-16}{(\xi-1)^{9}}$& $\frac{2}{(\xi-1)^{8}}$&$\frac{15}{12(\xi-1)^{7}}$ & $\frac{-5}{24(\xi-1)^{6}}$&  $\frac{-5}{288(\xi-1)^{5}}$& $\,\,\frac{28}{2400(\xi-1)^{4}}$&  \\
\hline\hline
$m_{2}^{\text{(p)}}=\sum\limits_{s=0}^{\infty} m_{2}^{\text{(p)}(s)}$
      & $m_{2}^{\text{(p)}(0)}$&  $m_{2}^{\text{(p)}(1)}$&  $\,\,m_{2}^{\text{(p)}(2)}$&  $\,\,m_{2}^{\text{(p)}(3)}$&  $\,\,m_{2}^{\text{(p)}(4)}$&  $\quad m_{2}^{\text{(p)}(5)}$&\strut\\
\hline
\hline
\end{tabular}
\end{center}
\vspace{-3mm}
\end{table}

The non-pole terms of ${\Large\textsl{m}}_{k}^{\text{(np)}} (\xi )$ can be calculated using two methods: either within the respective  tables (like table~\ref{tbl-smp1}), performing a partial summing over $s$ at fixed $k$ and $n$, or, which is easier, using the  obvious formulae
\begin{equation} \label{GrindEQ__30_}
{\Large\textsl{m}}_{2k}^{\text{(np)}} (\xi )={\Large\textsl{m}}_{2k}^{} (\xi )-{\Large\textsl{m}}_{2k}^{\text{(p)}} (\xi ),\qquad {\Large\textsl{m}}_{k,2k+n}^{\gtrless\text{(np)}} (\xi )={\Large\textsl{m}}_{k,2k+n}^{\gtrless}(\xi )-{\Large\textsl{m}}_{k,2k+n}^{\gtrless\text{(p)}} (\xi ).
\end{equation}
For example, the first three terms of non-pole MO at $k=1$ are the following ones
\begin{equation} \label{GrindEQ__31_}
{\Large\textsl{m}}_{2}^{\text{(np)}} (\xi )=-\frac{\alpha \nu [\alpha (1+\nu )]}{(\xi +1)^{2} } \,,
\end{equation}

\begin{equation} \label{GrindEQ__32_}
{\Large\textsl{m}}_{1,2+1}^{\text{(np)}>} (\xi )=\frac{28+20\xi+37\xi^{2}+6\xi^{3}-3\xi^{4}}{4(\xi+1)^{3}(\xi-2)^{2}}\alpha\nu[\alpha(1+\nu)]^{2}  ,
\end{equation}

\begin{equation} \label{GrindEQ__33_}
{\Large\textsl{m}}_{1,2+1}^{\text{(np)}<} (\xi )=\frac{-124+68\xi+15\xi^{2}+8\xi^{3}+\xi^{4}}{12(\xi+1)^{3}(\xi+2)^{2}}(\alpha \nu)^{2} [\alpha (1+\nu )] .
\end{equation}
The rest terms of ${\Large\textsl{m}}_{k,2k+n}^{\text{(np)}} (\xi )$ (at $k\leqslant k_{\text{max}}, n_{\text{max}}$) are easily obtained using the formula \eqref{GrindEQ__30_}, but we do not present them due to their sophisticated analytical form.

The developed theory defines a complete MO ${\Large\textsl{m}}(\xi )$, which is a real function of  $\xi $. Thus, the poles of g($\xi $) function \eqref{GrindEQ__4_} are found from the equation
\begin{equation} \label{GrindEQ__34_}
\xi={\Large\textsl{m}}(\xi ),
\end{equation}
which gives a renormalized spectrum of the system in the range ($-2<\xi<2$).

\section{Analysis of quasi-particle spectrum renormalized due to the multi-phonon processes in the vicinity of its main state at finite temperature}

The theory developed in the previous section makes it possible to calculate the renormalized energy spectrum of localized quasiparticle interacting with polarization phonons at $T\neq0$~K in a wide range of energies taking into account both unmixed and mixed multi-phonon processes. The proposed mathematical approach of MO analytical calculation differs from the one presented in \cite{Tka16}, where the MO terms describing the unmixed processes [${\Large\textsl{m}}_{\text u} (\xi )$] were completely taken into account using an exact partial summing, while the mixed ones [${\Large\textsl{m}}_{\text m} (\xi )$] --- in the first three orders over the coupling constant only. The ``new'', with respect to the simplified model \cite{Dav76}, satellite bound states of the system were revealed in \cite{Tka16}. However, further, in this paper we should show that some properties of their energy spectrum turned out to be an artifact of the approximated MO [${\Large\textsl{m}}_{\text m} (\xi )$] describing the mixed processes.

The renormalized energies of the main state ($e_{0} $) and complexes of bound states ($e_{-1}^{<\, 1}, e_{-1}^{>1}, e_{+1}^{<\, 1},$ $e_{+1}, e_{+1}^{>1},$ $e_{+1}^{>2} $) in the range ($-2<\xi<2$) were obtained from dispersion equation~\eqref{GrindEQ__34_}. In order to describe its properties and establish the mechanisms of formation, we calculated MO and its main terms as functions of $\xi $ at a different coupling constant ($\alpha $) and average phonons occupation number [$\nu $(T)].

\begin{figure}[!ht]
\centerline{\includegraphics[width=0.77\textwidth]{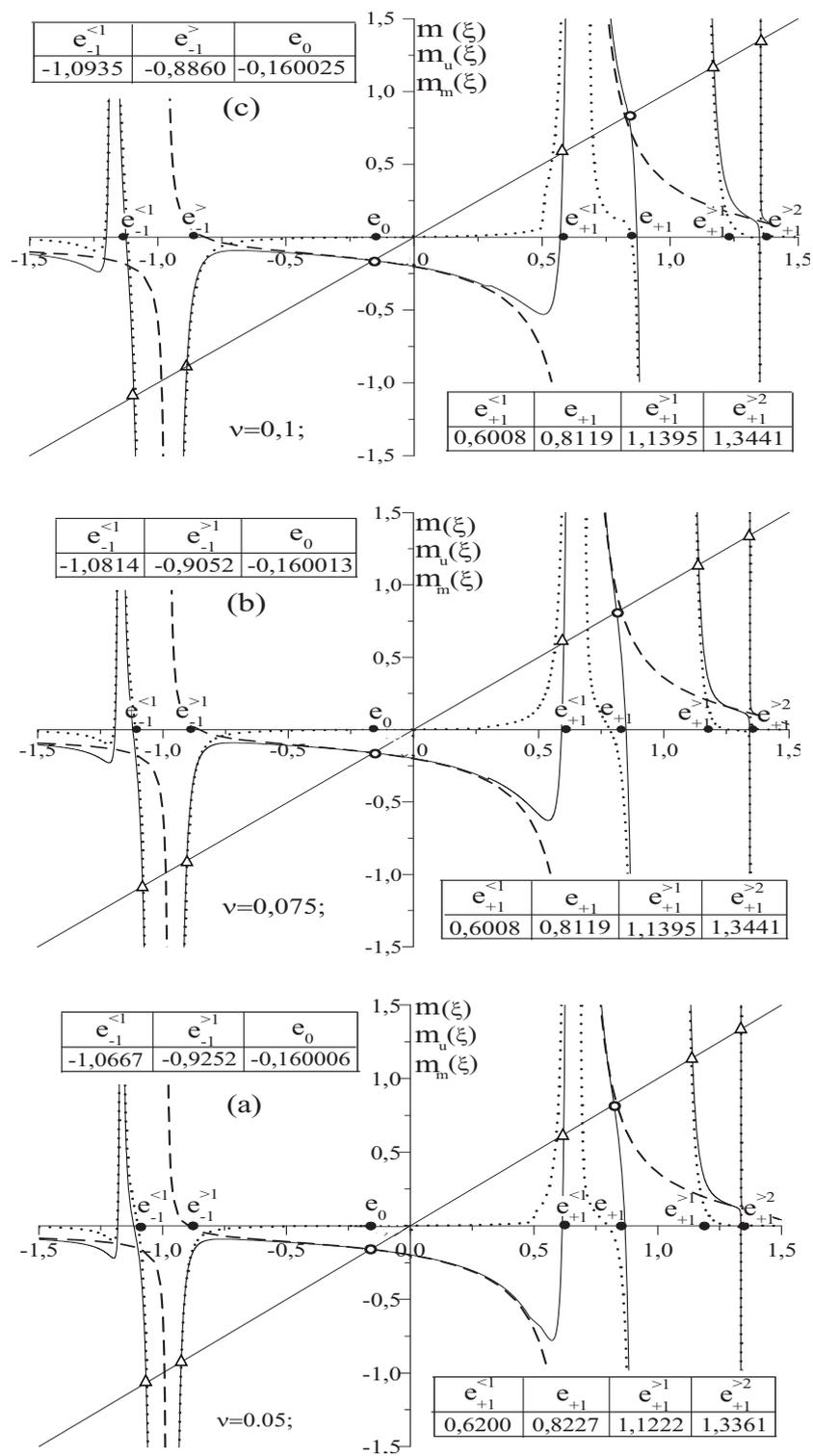}}
\caption{\label{fig2} MO ${\Large\textsl{m}}(\xi)$  with its terms ${\Large\textsl{m}}_{\text u}(\xi)$, ${\Large\textsl{m}}_{\text m}(\xi)$ and energy spectrum ($e_{-1}^{<1},\ldots, e_{+1}^{>2}$) as functions of~$\nu$ at $\alpha=0.16$.}
\end{figure}

In  figure~\ref{fig2}, the typical  picture of temperature  dependence of the  renormalized spectrum ($e_{-1}^{<\, 1},\ldots,e_{+1}^{>2} $) at $\alpha =0.16$, corresponding to the electron-phonon coupling with the energy $\sqrt{\alpha } \Omega =0.4\Omega $ is shown.  The dependences of MO ${\Large\textsl{m}}(\xi )$ and its main partial terms ${\Large\textsl{m}}_{\text u} (\xi )$ and ${\Large\textsl{m}}_{\text m} (\xi )$ are presented too, in order to reveal the role of unmixed and mixed processes in the formation of the respective energy levels of the system. A more detail contribution of different multi-phonon processes into the formation of the spectrum, for instance, is shown in table~\ref{tbl-smp3}.

Analysis of figure~\ref{fig2} and table~\ref{tbl-smp3} gives a clear understanding of the formation of the spectrum of a localized quasiparticle renormalized due to multi-phonon processes in the regime of a weak coupling at a finite temperature ($\nu\neq0$) and in the range of energies ($-2<\xi<2$). Figure~\ref{fig2}~(a) and table~\ref{tbl-smp3} prove that at $\alpha =0.16$, $\nu =0.05$, the two stationary states, i.e., the ground one with the energy $e_{0} $ and its first high-energy repeat ($e_{+1}$) are almost completely formed by the unmixed  [${\Large\textsl{m}}_{\text u} (\xi )\gg {\Large\textsl{m}}_{\text m} (\xi )$] processes of quasi-particle-phonon interaction with the prevailing creation of phonons [${\Large\textsl{m}}_{\text u}^{>} (\xi )\gg{\Large\textsl{m}}_{\text u}^{<} (\xi )$]. These energies are shifted into the low-energy region by the magnitude $\sim\alpha (1+\nu$), so that the difference between them ($\Delta e_{+1} =e_{+1} -e_{0} =0.982742$) is close to the energy of one phonon.

\begin{table}[!b]
\caption{Energy levels of the system and contributions of MO terms into their formation at $\alpha=0.16$; $\nu=0.05$.}
\vspace{2ex}
\label{tbl-smp3}
\begin{center}
\begin{tabular}{|p{0.2in}|p{0.5in}|p{0.5in}|p{0.5in}|p{0.5in}|p{0.5in}|p{0.5in}|p{0.5in}|p{0.5in}|} \hline \hline
  $e\backslash m$ & $\quad m$ & $\quad m_{\text u} $ & $\quad m_{\text u}^{>} $ &\quad $m_{\text u}^{<} $ & \quad $m_{\text m}^{} $ & $\quad m_{\text m}^{>} $ & $\quad m_{\text m}^{<} $ & $ \quad m_{e}^{} $ \\ \hline \hline
$e_{+1}^{>2} $ & 1.33606 & 0.11892 & 0.11548 & 0.00343 & 1.20826 & 1.20802 & 0.00025 &0.00889 \\ \hline
$e_{+1}^{>1} $ & 1.12225 & 0.23894 & 0.23516 & 0.00378 & 0.88331 & 0.88244 & 0.00087 & 0.02120 \\ \hline
$e_{+1}^{} $ & 0.82274 & 0.89616 & 0.89175 & 0.00440 & $-$0.07342 & $-$0.07202 & $-$0.00139 & 0.00187 \\ \hline
$e_{+1}^{<\, 1} $ & 0.62004 & $-$1.92236 & $-$1.92732 & 0.00496 & 2.54237 & 2.54255 & $-$0.00018 & 0.00335 \\ \hline 
$e_{0}^{} $ & $-$0.16001 & $-$0.15985 & $-$0.16948 & 0.00962 & $-$0.00015 & $-$0.00009 & $-$0.00096 & $-$0.00089 \\ \hline 
$e_{-1}^{>1} $ & $-$0.92518 & 0.01159 & $-$0.09394 & 0.10553 & $-$0.96593 & $-$0.87066 & $-$0.09527 & 0.02949 \\ \hline
$e_{-1}^{<\, 1} $ & $-$1.06672 & $-$0.16821 & $-$0.08549 & $-$0.08272 & $-$0.88554 & $-$0.66469 & $-$0.22085 & 0.00692 \\ \hline \hline
\end{tabular}
\end{center}
\end{table}

In the high-energy region, one can see  three stationary bound states, which are satellite to the first main one: the lower one with the energy $e_{+1}^{<\, 1} $, smaller than $e_{+1} $, and two upper states with the energies $e_{+1}^{>1}, e_{+1}^{>2} $ bigger than $e_{+1} $. These satellite states are mainly formed by the mixed [${\Large\textsl{m}}_{\text m} (\xi )\geqslant {\Large\textsl{m}}_{\text u} (\xi )$] processes of quasi-particle-phonon interaction with the prevailing creation of phonons [${\Large\textsl{m}}_{\text m}^{>} (\xi )\gg{\Large\textsl{m}}_{\text m}^{<} (\xi )$]. We should note that the lower satellite state with the energy $e_{+1}^{<\, 1} $ is also a stationary one, taking into account the multi-phonon processes.

The state with the energy ($ e_{+1}^{<\, 1} $) does not degenerate with the main one with the energy ($e_{+1} $)  and, thus, it does not decay  contrary to the results of \cite{Tka16}, where the positive contributions only of two- and three-phonon mixed processes (${\Large\textsl{m}}_{\text m} $) were taken into account. It was not sufficient to compensate the negative contributions produced by unmixed processes (${\Large\textsl{m}}_{\text u} $) into the formation of both energy levels when $\nu $ increased. Now, from  table~\ref{tbl-smp3} one can clearly see that, among all high-energy levels, only the one with $e_{+1}^{<\, 1} =0.62$ is formed by comparable but opposite sign terms of MO ${\Large\textsl{m}}_{\text m} \approx {\Large\textsl{m}}_{\text m}^{>} =2.54$ and ${\Large\textsl{m}}_{\text u} \approx {\Large\textsl{m}}_{\text u}^{>} =-1.92$. Thus, it is necessary to take into account the multi-phonon processes in order to correctly calculate ${\Large\textsl{m}}_{\text m} $, the same as for ${\Large\textsl{m}}_{\text u} $.

Figure~\ref{fig2} and table~\ref{tbl-smp3} prove that the formation of satellite high- and low-energy states is similar. The energies of low-energy states, which are denoted as ($e_{-1}^{<\, 1}, e_{-1}^{>1} $), respectively, are also mainly produced by the contributions of mixed multi-phonon processes [$\left|{\Large\textsl{m}}_{\text m} (\xi )\right|\gg\left|{\Large\textsl{m}}_{\text u} (\xi )\right|$]. Herein, the contributions of the processes accompanied by the creation of phonons are bigger than that of annihilation [$|{\Large\textsl{m}}_{\text m}^{>} (\xi )|\gg|{\Large\textsl{m}}_{\text m}^{<} (\xi )|$]. The both satellite low-energy states are the stationary.

\begin{figure}[!t]
\centerline{\includegraphics[width=0.95\textwidth]{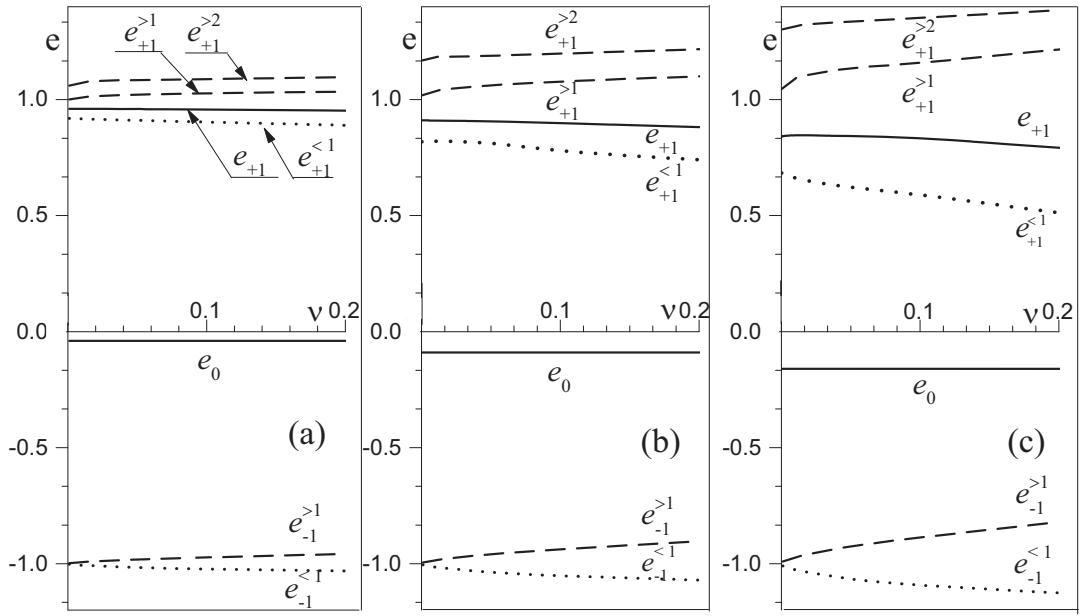}}
\caption{\label{fig3} Energy levels as functions of the average phonon occupation number ($\nu$) at $\alpha=0.04$~(a); $\alpha=0.09$~(b); $\alpha=0.16$~(c). }
\end{figure}

Temperature dependences of renormalized energies of the system at $\alpha=0.04, 0.09, 0.16$, which correspond to the magnitude of interaction 0.2, 0.3 and 0.4, respectively, (in units of phonon energy $\Omega $) are shown in figure~\ref{fig3}. It is clear that in the regime of weak coupling, the temperature dependence of the spectrum is qualitatively similar. When $\nu $ (temperature) increases, the upper satellite low-and high-energy levels ($e_{-1}^{>1},e_{+1}^{>1}, e_{+1}^{>2}$) smoothly (quasi-linearly) shift into a high-energy region while the main level ($e_{0} $), its first phonon repeat ($e_{+1} $) and both low-energy levels ($e_{-1}^{<\, 1}$, $e_{+1}^{<\, 1} $ ) smoothly shift into the low-energy region of the spectrum. The magnitudes of the shifts of all levels slightly increase at a bigger $\alpha $.

\section{Main results and conclusions}
The main result of this paper is that the theory of renormalized energy spectrum of a localized quasi-particle weakly interacting with polarization phonons at a finite temperature is developed within the Feynman-Pines diagram technique using the Frohlich model and taking into account the multi-phonon processes. Now it is possible to obtain the renormalized energies of the main state and the complexes of bound states.

The created computer program exactly calculates the analytical expressions for the diagrams of MO in an arbitrary order over the coupling constant, which is limited by a computer resource only. The analysis of the diagrams till the tenth order made it possible to separate its pole and non-pole terms and perform a partial summing of their main terms. The analytical expressions for the general terms in the ranges of main diagrams are obtained and partially summed. As a result, a complete MO is obtained in the range $-2<\xi<2$ taking into account the multi-phonon processes.

It is shown that a sufficient consideration of multi-phonon processes in MO provides a stationary energy spectrum of the system. Its properties are similar, in certain areas of the spectrum, and somewhere differ from that obtained both in the model with an additional condition for the quasi-particles operators \cite{Dav76} and in the Frohlich model, recently observed in \cite{Tka16}, where the unmixed processes were accounted exactly while the mixed ones --- only in the second and third orders of MO diagrams. It is established that, like in  \cite{Tka16},  contrary to the model used in \cite{Dav76}, the spectrum of the system is not equidistant and contains  ``new'' complexes of bound states. The latter are stationary and do not decay in a high-energy region as in  \cite{Tka16} due to the above mentioned approximation in the mixed MO.

The renormalized energy of the main state and its high-energy phonon repeat are mainly formed by interaction accompanied by the creation of phonons in unmixed processes, while all satellite low- and high-energy states are mainly formed by the interaction accompanied by  annihilation of phonons in mixed processes.

It is shown that in the regime of a weak coupling, when temperature increases, the upper satellite low-and high-energy levels ($e_{-1}^{>1}, e_{+1}^{>1}, e_{+1}^{>2} $) smoothly quasi-linearly shift into a high-energy region while the main level ($e_{0} $), its first phonon repeat ($e_{+1} $) and low-energy levels ($e_{-1}^{<\, 1}, e_{+1}^{<\,1} $) smoothly shift into the low-energy region of the spectrum.

The final conclusion is that using a computer program for  analytical calculation of the high orders of MO diagrams, we effectively took into account the multi-phonon processes of interaction between the quasi-particle and polarization phonons, thus essentially clarifying the properties of the complexes of bound states near the main one. The developed theoretical approach together with a proper computer support can be successfully used for the calculation of the renormalized spectrum of this system in a wider range of energy and coupling constant. It can be used for the solution of the other problems where the quasi-particles interacting with phonons are investigated.

\section*{Acknowledgements}
The authors would like to thank V.A. Tarnovetsky for the help in analytical programming of Feynman-Pines diagrams.

\appendix
\section{Algorithm of diagrams calculation} \label{App1}

The algorithm is based on a brute-force search of all possible variants of diagrams. Each diagram is converted into its analytic representation and  then they all are aggregated into a single analytic formula.

To this end,  several steps should be done:

1. Generate all possible dashed lines directions of each diagram.

2. Generate all possible diagram shapes with filtering out the diagrams with non-covered segments.

3. Convert each diagram into an analytic representation and aggregate all of them into a single formula.

\subsection{All possible diagrams}

Generation of all diagrams for some power on $\nu$ is implemented as a recursive function with depth~$N$. The shape of the diagram is encored with a list of $N$ number pairs where every pair  encodes the start and end point of a single line. The algorithm starts from an empty list and uses recursive calls to populate the list with all possible variations.

We shall use $N=3$ in the examples. The list is initially empty: [ ].

\begin{figure}[!ht]
\centerline{\includegraphics[width=0.28\textwidth]{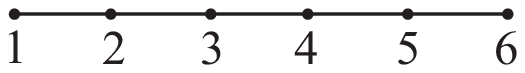}}
\end{figure}

On each recursive call we select the first available point as a starting point and enumerate all points on its right  as the end points.

\begin{figure}[!ht]
\centerline{\includegraphics[width=0.28\textwidth]{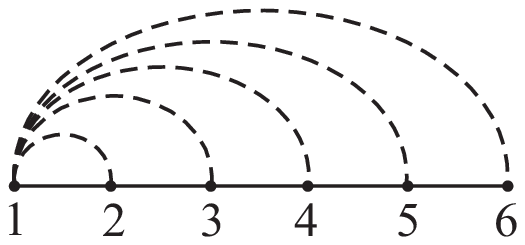}}
\end{figure}

Corresponding lists are: [(1,2)], [(1,3)], [(1,4)],	[(1,5)], [(1,6)].

For each of $2N-1$ variation, we call the function recursively again and again, and it selects the first available point as a starting point and enumerates all possible ends:

\begin{figure}[!ht]
\centerline{\includegraphics[width=0.28\textwidth]{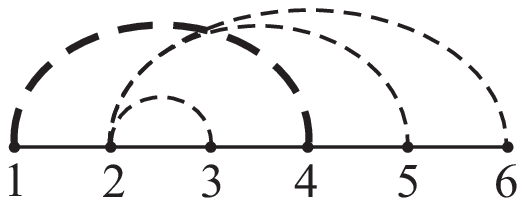}}
\end{figure}

Corresponding lists are: [\textbf{(1,4)}, (2,3)], [\textbf{(1,4)}, (2,5)],	[\textbf{(1,4)}, (2,6)].

Bold dashed line is a line already fixed earlier and thin dashed lines show all possible variations of lines in the current call.

For every variation received from the first level we shall have $2(N-1)-1$ variations at the second level and for each of them, as before, we call the function recursively again.

At the third level in our $N=3$ example, we shall always have only one possible option.

\begin{figure}[!ht]
\centerline{\includegraphics[width=0.28\textwidth]{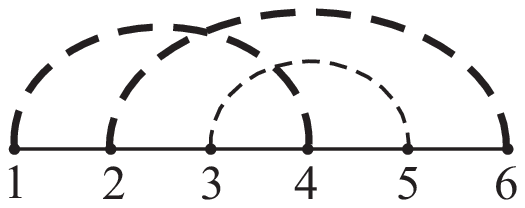}}
\end{figure}
A corresponding list is [\textbf{(1,4)}, \textbf{(2,6)}, \textbf{(3,5)}].

Finally, once we have fixed this single possible option, we have fully finished the diagram for a given~$N$.

In total, the third level of the recursive call will be called $2N-1 + 2(N-1) - 2(N-2) + \ldots 1$ times giving us all possible and unique shapes of the diagrams.

\subsection{Filtering out non-fully covered diagrams}

Non-fully covered diagrams are not of interest to us and we just neglect them:

\begin{figure}[!ht]
\centerline{\includegraphics[width=0.28\textwidth]{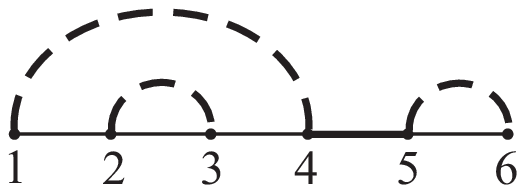}}
\end{figure}

\subsection{All possible directions}

For each fully covered diagram, we enumerate all possible  directions of dashed lines. This is implemented as a single loop from 0 to $2^N-1$. Each number in binary representation represents one possible variation of the lines direction where 0 means the left-hand direction and 1 means right-hand direction. In our example of $N=3$, we iterate from 0 to 7 where:

a. Binary representation of 0 is 000 which means all lines are directed to the left.

b. Binary representation of 7 is 111 which means all lines are directed to the right.

c. Binary representation of 3 is 011, i.e., 1st digit is one, 2nd digit is 1 and the this digit is 0. This encodes one of the diagrams above as:
[(1,4, right-hand), (2,6, right-hand), (3,5, left-hand)].
\begin{figure}[!ht]
\centerline{\includegraphics[width=0.28\textwidth]{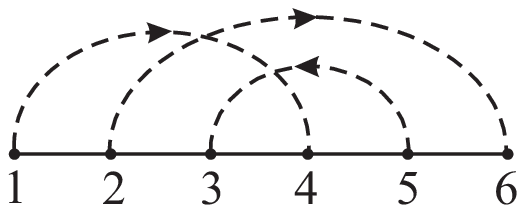}}
\end{figure}

\section{Functions of MO's of higher orders}\label{App2}

\begin{equation}
\varphi_{1,3}^{>} (\xi)=2\{-7+\xi[-5+\xi (-19+15\xi)]\},
\end{equation}
\begin{equation}
\varphi_{1,4}^{>}(\xi)=4[-93+\xi(-378+\xi\{220+\xi[-196+\xi(567-386\xi +74\xi ^{2})]\})],
\end{equation}
\begin{align}
&\varphi_{1,5}^{>} (\xi )=2(48768+\xi (78600+\xi (87291+\xi (-434688+\xi (406354+\xi(-345016
\nonumber\\
&+\xi (394832+\xi (-294448+\xi (116222+\xi (-22816+1765\xi )\ldots),
\end{align}
\begin{align}
&\varphi_{1,6}^{>} (\xi )=4(27145440+\xi (66845040\xi (-109082040+\xi (46249494+\xi (-336576266
\nonumber\\
&+\xi (812355829+\xi (-932558268 +\xi (781847971+\xi (-622684844+\xi (444206106
\nonumber\\
&+\xi (-238988488+\xi (88941524+\xi (-22008858+\xi (3453473+7\xi(-44428+1749\xi )\ldots),
\end{align}
\begin{align}
&\varphi_{2,5}^{>} (\xi)=4(-37824+\xi(-5024+\xi (-54576+\xi (217832+\xi (120172+\xi(11584
\nonumber\\
&+\xi (76889+ \xi(-132534+\xi (-32193+\xi (60604+\xi (-2329+\xi (-8782+1765\xi)\ldots),
\end{align}
\begin{align}
&\varphi_{2,6}^{>} (\xi )=2(-87229440+\xi (-709003776+\xi (0853504+\xi (-105895132+\xi (2839877360
\nonumber\\
&+\xi (-132092208+\xi (211544419+\xi (- 280680956+ \xi(-1856504944+\xi (545314558
\nonumber\\
&+\xi(-30531013+\xi(178915528+\xi(2277094+\xi (85013000+\xi (-25160882
\nonumber\\
&+\xi (-6144178+\xi(4603407+35\xi (-25712+1749\xi )\ldots),
\end{align}

\begin{equation}
\varphi_{2}^{(\text e)}(\xi)=4\xi,
\end{equation}
\begin{equation}
\varphi_{4}^{(\text e)}(\xi )=12\xi (212+135\xi ^{2} -168\xi ^{4} +37\xi ^{6}),
\end{equation}
\begin{align}
&\varphi_{6}^{(\text e)}(\xi )=8\xi (-316439136-742497696\xi ^{2}+1198156746\xi ^{4} -459531015\xi ^{6}-85177928\xi ^{8}
\nonumber\\
&+122619187\xi ^{10}-42134414\xi ^{12} +7076639\xi ^{14} -597748\xi ^{16}+20405\xi ^{18}).
\end{align}

\vspace{-3mm}
\ukrainianpart

\title{Перенормування енергетичного спектру локалізованих квазічастинок багатофононними процесами при скінченних температурах}
\author{М.В.Ткач, О.Ю.Питюк, О.М.Войцехівська, Ю.О.Сеті}
\address{Чернівецький національний університет ім.~Ю.~Федьковича, \\вул. Коцюбинського, 2,  58012 Чернівці, Україна}

\makeukrtitle

\begin{abstract}
\tolerance=3000%
На основі діаграмної техніки Феймана-Пайнса запропонована теорія перенормування енергетичного спектру локалізованої квазічастинки, яка взаємодіє з поляризаційними фононами при скінченній температурі.
Розроблена комп'ютерна програма, яка ефективно враховує багатофононні процеси, точно визначаючи усі діаграми масового оператора і їх аналітичні вирази у довільному порядку  за константою зв'язку. Це дозволило розмежувати неполюсні й полюсні доданки масового оператора та виконати парціальне підсумовування їх головних складових.
Розв'язанням дисперсійного рівняння отримано перенормований спектр системи в околі основного стану, де спостерігаються високо- та низькоенергетичні комплекси зв'язаних станів. Проаналізовано властивості спектру в залежності  від константи зв'язку й температури.
\keywords квазічастинка, фонон, функція Гріна, масовий оператор

\end{abstract}

\end{document}